\documentclass[12pt]{iopart}
\usepackage{graphicx}

\begin{document}

\title[The role of oxygen and surface reactions in the deposition of SiO$_x$ films]{The role of oxygen and surface reactions in the deposition of silicon oxide like films from HMDSO at atmospheric pressure}

\author{R. Reuter, K. R\"ugner, D. Ellerweg, T. de los Arcos, A. von Keudell, J. Benedikt}
\address{Research Department Plasmas with Complex Interactions, Ruhr-Universit\"at Bochum,
  Universit\"atsstr. 150, 44780 Bochum, Germany}

\begin{abstract}

The deposition of thin SiO$_x$C$_y$H$_z$ or SiO$_x$H$_y$ films by means of atmospheric pressure
microplasma jets with admixture of Hexamethyldisiloxane (HMDSO) and oxygen and the role of surface
reactions in film growth are investigated. Two types of microplasma jets, one with a planar
electrodes and operated in helium gas and the other one with a coaxial geometry operated in argon,
are used to study the deposition process. The growth rate of the film and the carbon-content in the film are
measured as a function of the O$_2$ and HMDSO admixture in the planar jet and are compared to mass
spectrometry measurements of the consumption of HMDSO. Additionally, the localized nature of the
jet-substrate interaction is utilized to study surface reactions by applying two jets on a rotating
substrate. The addition of oxygen into the gas mixture increases HMDSO depletion and the growth
rate and results in the deposition of carbon free films. The surface reaction is responsible for
the carbon removal from the growing film. Moreover, carbon free films can be deposited even without
addition of oxygen, when coaxial jet operated with argon is used for the surface treatment. We
hypothesize that ions or excited species (metastables) could be responsible for the observed
effect.

\end{abstract}

\maketitle

\section{Introduction}

Silicon dioxide (SiO$_2$) is a widely used thin film material. It is the most common dielectric in semiconductor technology, serves as corrosion protection or permeation barrier in the packaging industry or is used as a scratch resistant coating on polymers. SiO$_x$ films are usually deposited by plasma enhanced chemical vapour deposition at low pressure and its deposition has been extensively studied in the past\cite{Magni2001, Kim1997a, Ricci2010a, Leu2003, Aumaille2000, Hegemann1999a, Hegemann2003}. Attractive alternative to these low pressure processes is the SiO$_x$ deposition at atmospheric pressure, where no vacuum systems and batch processing would be necessary in a production line. Several approaches to obtain SiO$_x$ at atmospheric pressure have been studied\cite{sonnenfeld02, Starostine07, Enache07, Alexandrov05, Massines05, Premkumar2009a, Martin2004a, Sawada1995, Fanelli2010b}. One of these possible systems are microplasma jets that operate at low powers (\textless 10W) and allow a localized deposition of these films. Usually, a small amount of precursors like Hexamethyldisiloxane (HMDSO) and oxygen are admixed into a main noble gas flow (He or Ar) and SiO$_2$ films are deposited\cite{Raballand2008b, Raballand2009a, Schafer2008, Lommatzsch09}. Many investigations regarding the deposition of SiO$_x$ films have been carried out in our group\cite{Raballand2008b, Raballand2009a} in the past. A coaxial jet driven in argon with admixtures of HMDSO and O$_2$ has been used to deposit SiO$_x$C$_y$H$_z$ films. It has been found that even without addition of oxygen this coaxial jet is able to deposit carbon free films. However, the plasma chemistry leading from HMDSO/O$_2$ gas mixtures to SiO$_2$ films is not very well understood yet. The main goal of this work is to understand this chemistry without concentrating much on the improvement of film quality or deposition rate. We will show that microplasma jets are well suitable for this study since they allow measurements of gas phase plasma chemistry products at the deposition area and spatial separation of the HMDSO based deposition and O$_2$ based surface treatment.

\section{Experimental setup}

Two different microplasma jets, a planar jet with He as plasma forming gas and a coaxial
microplasma jet with Ar as plasma forming gas are used in this study. The geometry of the planar
microplasma jet used in this study is based on the microplasma jet design optimized for good optical
access for plasma diagnostics\cite{Schulz-vonderGathen2007, Knake2008}. The geometry of this planar
jet was slightly changed to adopt it for the deposition process with HMDSO: The channel cross
section is the same as in the previous studies, 1x1\,mm$^2$, but the electrode length was reduced
to 10\,mm and the gas flow was increased to 5\,slm of He to maintain a short residence time of
HMDSO in the jet. The possible formation of dust particles is avoided and the deposition on the
electrode surface in the jet is reduced. Additionally, the residence time of the species in the jet $\tau_{res} = l_{jet} \cdot v_{gas}^{-1} = 1,2 \cdot 10^{-4}s$, where $l_{jet}$ is the length of the discharge channel and
$v_{gas}$ the gas velocity in the jet, is in this case much shorter than
the diffusion time $\tau_{diff} = \Lambda_{diff}^2 \cdot D_0^{-1} = 8.6 \cdot 10^{-4}s$ with
assuming $\Lambda_{diff} = r/2.405$ and $D_0 = 5 \cdot 10^{-5}\,m^2s^{-1} $. This has the
consequence that diffusion losses of reactive gas phase species are negligibly small and that, the
other way around, the processes on the wall of the jet have a negligible effect on the deposition
at the substrate, because particles that originate from the inner surface of the plasma jet are
unable to diffuse into the central region of the plasma and cannot therefore reach the central
point of impact of the effluent on the substrate.

\begin{figure} \centering
\includegraphics*[width=8cm]{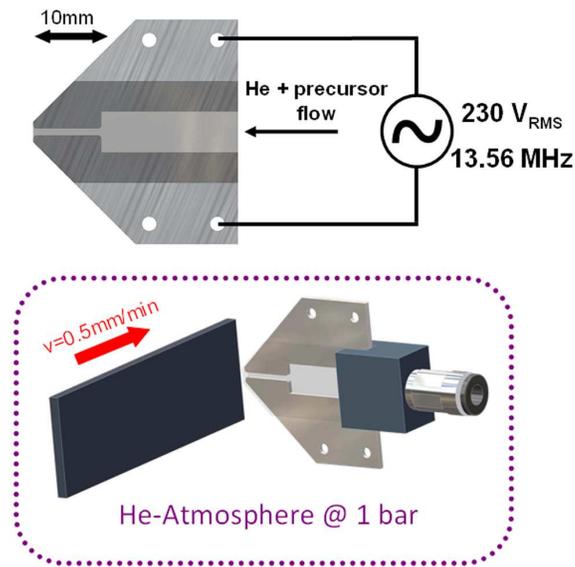}
\caption{Planar jet and the linear drive process}
\label{fig:fig1}
\end{figure}

One Electrode is powered with a frequency of 13.56\,MHz and a voltage of 230\,V$_{RMS}$, the other
electrode is grounded. More detailed description of the experimental setup can be found
elsewhere\cite{Reuter2011}. This jet was designed to study solely the plasma chemistry and the
deposition process. Its operation with He and He/O$_2$ has been studied in the past\cite{Schulz-vonderGathen2007, Knake2008}. The plasma is localized in the region
between the electrodes without direct contact with the treated surface, which is placed at 4\,mm
distance from the jet nozzle. The lifetime of the jet is limited due to deposition on the inner
wall, when the jet is operated with admixtures of HMDSO. The lifetime was determined by running the
jet with admixtures of HMDSO (0.1\,sccm) and O$_2$ (2\,sccm) and moving it in front of a substrate
with a constant velocity for five hours. Afterwards the film thickness was measured at various
locations along the deposited film track. It was observed that the deposition rate strongly
decreases after a deposition time of around four hours. In order to ensure constant deposition
conditions, the jet was changed every two hours.

\begin{figure} \centering
\includegraphics*[width=8cm]{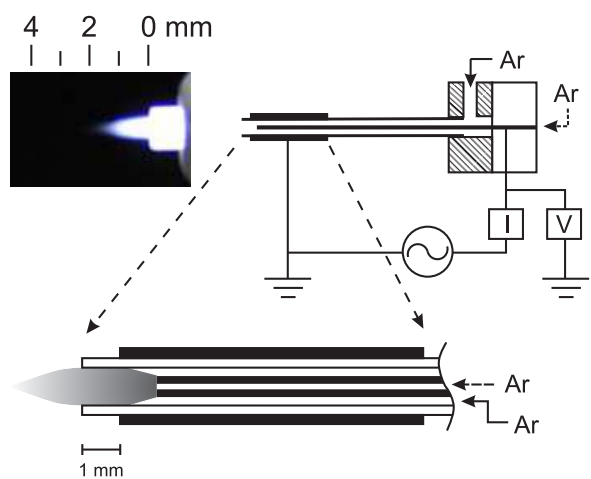}
\caption{The scheme of the coaxial jet with the photograph of the plasma at the end of the ceramic
tube} \label{fig:fig2a}
\end{figure}

The second plasma source, the coaxial microplasma jet, consists of a stainless steel capillary
inserted into a ceramic tube. An annular space with gap of 250\,$\mu$m between the tube and the
capillary is formed in this way. An aluminium tube outside the ceramic tube serves as grounded
counter electrode. The capillary is powered with a frequency of 13.56\,MHz and a voltage of
230\,V$_{RMS}$. The jet is driven with argon with an inner flow through the capillary (160\,sccm)
and an outer flow between capillary and ceramic tube (3000\,sccm) \cite{Benedikt2007c}. Argon
plasma is formed at the end of the capillary and it extends few mm from the jet, cf.
Figure~\ref{fig:fig2a}. A detailed description of the geometry and experimental results can be
found elsewhere\cite{Raballand2009a, Benedikt2006, Benedikt2010}. The deposition was typically
performed on the substrate at 1\,mm distance from the jet, where the plasma is in direct contact
with it. As already mentioned before, carbon free SiO$_2$ films can be deposited with this jet
without addition of O$_2$. Moreover, the jet is designed and the gas flows chosen in such a way
that deposition inside the jet is minimized and the lifetime of this jet is therefore much longer.

The HMDSO liquid injected into the jet was vaporized by a bubbling system at a temperature of
0\,$^{\circ}$C. Each deposition was performed in a helium atmosphere at atmospheric pressure in a
small chamber (volume of 5\,liters) to ensure controlled deposition conditions in each experiment.
It has to be kept in mind that the HMDSO gas diffuses into the whole volume of the chamber leading
to a HMDSO concentration in the helium atmosphere of around 20\,ppm. Additionally, some impurities
are very probably present in this atmosphere due to desorption of gases from the reactor wall.

\begin{figure} \centering
\includegraphics*[width=8cm]{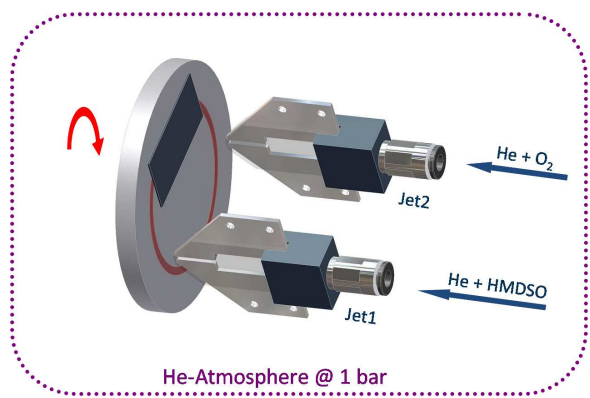}
\caption{Rotating disc process - Two planar jets in parallel in front of a rotating disc with mounted substrate and red deposition track.}
\label{fig:fig2}
\end{figure}

Crystalline silicon wafers are used as a substrate for the deposition and two different deposition
schemes are used. Only one plasma jet was used in the first one. 15x30\,mm$^2$ silicon substrates
were located 4\,mm in front of the jet as indicated in Figure~\ref{fig:fig1}. To achieve a nearly
homogeneous deposition in one direction, the substrate was moved in front of the jet with a
velocity of 0.5\,mm$\cdot$min$^{-1}$. As will be shown later, a line with a bell-shaped profile and
with full width of half maximum (FWHM) of about 4 mm is deposited. We call this deposition process
``linear drive'' process.
Two jets are simultaneously used in the second deposition scheme in such a way that the surface
treatment takes place on the same track of a rotating substrate, see Figure~\ref{fig:fig2}. The setup
consists of two jets running in parallel with a distance of 40\,mm in between. A substrate holder
is rotated with a certain frequency in front of the jets. The distance between the jets and the
substrate mounted on the rotating substrate holder is 4\,mm.

An alternating process is realized by rotating the substrate, where the substrate is treated by the first jet and half a period later by the second jet. Rotating the substrate gives the opportunity to separate the deposition process into two independent steps. One is the deposition process with a HMDSO plasma without oxygen, the other one is the treatment of the surface by an oxygen plasma only. We call this deposition process ``rotating disc'' process. The rotation frequency was varied from 0.0006\,Hz up to 5\,Hz. A frequency of 0.06\,Hz leads to a deposition of one monolayer per rotation.


\section{Film diagnostics}

The relative C-content is measured by using Fourier-transformed Infrared Spectroscopy (FTIR).
Moving the substrate respectively the jet in the deposition process leads to a bell-shaped profile
perpendicular to the moving direction. To measure the film with the FTIR, the substrate is mounted
on a 1\,mm broad slit to ensure that the infrared light can only pass the film at the position of the
maximum film thickness. Transmission is measured and transformed into an absorption spectrum. The
baseline is fitted with polynomials and subtracted. SiO$_2$ exhibits three characteristic
absorption peaks, namely, rocking (450\,cm$^{-1}$), bending\,(800\,cm$^{-1}$), and asymmetrical
stretching [1075\,cm$^{-1}$ (AS1) and 1150\,cm$^{-1}$ (AS2)]\cite{Pai86}. AS1 and especially AS2
are relatively broad and overlap partially. If CH$_x$ is contained in the film, Si atoms are bonded
to CH$_x$ groups and the AS1 peak is slightly shifted to lower wavenumbers\cite{Borvon02}. The
Si(CH$_3$)$_x$ bending absorption peak at around 1260\,cm$^{-1}$, which shifts to higher
wavenumbers and decreases for lower carbon content, also indicates the presence of carbon in the
film. The shift of the AS1 peak and the presence of the Si(CH$_3$)$_x$ absorption can be used as a
qualitative measure for the carbon content. To compare qualitatively the carbon content, the peak area of the
Si(CH$_3$)$_x$ bending absorption peak at around 1260\,cm$^{-1}$ taken from the normalized FTIR spectrum was
used. Additionally, the presence of CH$_3$ rocking in Si(CH$_3$)$_2$ and Si(CH$_3$)$_3$ at 800\,cm$^{-1}$ and 840\,cm$^{-1}$ and C-H streching at 2960\,cm$^{-1}$ peaks shows a high carbon content\cite{Borvon02}. The very broad OH absorption peaks between 3000\,cm$^{-1}$ and 3750\,cm$^{-1}$ and at around 940\,cm$^{-1}$ indicates OH goups in the SiO$_x$H$_y$ film. The FTIR spectra were normalized to the highes value in the spectrum
(AS1 peak).

The quantitative film composition was investigated by x-ray photoelectron spectroscopy(XPS). XPS measurements have been carried out in a Versaprobe spectrometer (Physical Electronics, PHI 5000 VersaProbe) using monochromatic Al K$\alpha$ (1486.6\,eV) radiation. The measuremnt spot had a diameter of 100\,$\mu$m. This is quite small compared to the dimensions of the deposited film's profile (Fig ~\ref{fig:fig4}).
The deposited films were exposed to the ambient atmosphere for several days before XPS measurements. To test the effect of this exposure, depositions under the same conditions were
carried out inside the load lock of the XPS device allowing XPS analysis without exposure of films
to ambient atmosphere. Additionally, the top layer was sputtered before
measuring the film composition to enable a measurement of the film bulk material.

The film thickness (deposition rate) was measured by spectroscopic ellipsometry (M-88 rotating analyzer, J.A. Wollam
Co., Inc.). The diameter of the light beam was reduced to 1 mm and the measurements were performed
in the middle of the line deposited in the linear drive process. The film thickness is calculated
from the measured ellipsometric parameters $\Phi$ and $\Delta$ by applying an optical model that
consists of a thick Si substrate and a SiO$_2$ layer (even for carbon rich films) with thickness as
fit parameter. The thickness uniformity is also fitted and is typically around 5\%. An extra
roughness layer was neglected, because its addition into the model did not improve the fit.
Profilometer (Dektak 6m, Veeco) measurements were performed to validate the ellipsometry results. The profile perpendicular to the deposition direction is measured by measuring the step height after mechanically scratching off few mm broad region of the film perpendicular to the deposition direction. As will be shown later, the results of ellipsometer and profilometer measurements are in very good agreement.

The depletion of HMDSO precursor in the plasma was measured by means of molecular beam mass
spectrometry (MBMS). The change of the height of the mass peak at 147\,amu (the most intense peak
in the HMDSO spectrum) before and after plasma ignition were compared and used to calculate HMDSO
depletion. The MBMS setup was described in detail elsewhere\cite{Ellerweg2010a, Benedikt2009a}.


\section{Results and discussion}

This section is separated into three parts. In the first part, the deposition by a planar jet under
varying HMDSO and oxygen flows is described. The second passage deals with experiments carried out
on a rotating substrate. Surface reactions involved in the deposition process can be separated from
gas phase processes in this way. In the last passage, the deposition of carbon-free films by means
of coaxial jet with argon gas and without addition of oxygen is explored.

\subsection{Deposition by planar jet}


It has already been shown that the carbon content in the film is large without any addition of oxygen. An increasing oxygen flow leads then to a decrease of the carbon content\cite{Reuter2011}.
In the first experiment carried out in the linear drive process, the carbon content in the film was now studied as a function of the O$_2$ and HMDSO flows. The oxygen flow was varied over a wide range from 0\,sccm up to 20\,sccm. The HMDSO flow was chosen between 0.01\,sccm and 0.2\,sccm.

\begin{figure} \centering
\includegraphics*[width=8cm]{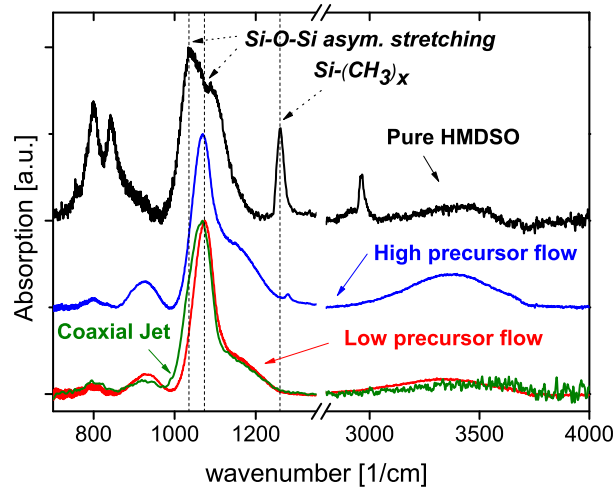}
\caption{Normalized FTIR spectra of film deposited in the linear drive process. (Conditions: Pure HMDSO: 0.1\,sccm (20\,ppm) HMDSO, 0\,sccm O$_2$; Low precursor flow: 0.01\,sccm (2\,ppm) HMDSO, 0.2\,sccm O$_2$; High precursor flow: 0.1\,sccm (20\,ppm) HMDSO, 2\,sccm O$_2$; Coaxial Jet: 0.02\,sccm (4\,ppm) HMDSO, 0\,sccm O$_2$ \cite{Raballand2009a}) The FTIR spectra for the conditions of pure HMDSO and of high precursor flow are shifted for a better visibility.}
\label{fig:fig3}
\end{figure}

Figure~\ref{fig:fig3} presents three FTIR spectra of films deposited with a single planar jet in
the linear drive process and one FTIR spectrum from a film deposited with the coaxial jet for
comparison. A film with a high carbon content is formed without any addition of oxygen as
identified by the high Si(CH$_3$)$_x$ bending absorption peak at around 1260\,cm$^{-1}$ and by the
shift of AS1 and AS2 to lower wavenumbers. The carbon content decreases when oxygen is added to the
gas mixture and its flow increases.

A reduction of the carbon content below the detection limit of the FTIR (measured by the presence
of the 1260\,cm$^{-1}$ peak in the FTIR spectrum) can, however, be achieved by decreasing the HMDSO
and O$_2$ flows (with the constant ratio O$_2$/HMDSO\,=\,20), see Figure~\ref{fig:fig3}. These
spectra are in good agreement with those measured by F. Massines et. al.\cite{Massines2005a}, who
deposited thin films from HMDSO by an atmospheric pressure Townsend dielectric barrier discharge in
mixtures of N$_2$O and HMDSO in N$_2$.

Carbon free films can also be deposited by the coaxial jet\cite{Raballand2009a}, which is
possible even without addition of O$_2$ into the gas mixture. The FTIR spectrum of such a film is
shown in Figure~\ref{fig:fig3}. The deposition of carbon-free films without addition of oxygen
using the planar jet has never been achieved, even with very low HMDSO flows (2\,ppm). There are
small differences in the FTIR spectra deposited with planar jet with He/HMDSO/O$_2$ gas mixture and
coaxial jet with Ar/HMDSO gas mixture. The width of the AS1 peak is smaller and the OH absorption
features are bigger for the material deposited by means of the planar jet.

\begin{figure} \centering
\includegraphics*[width=8cm]{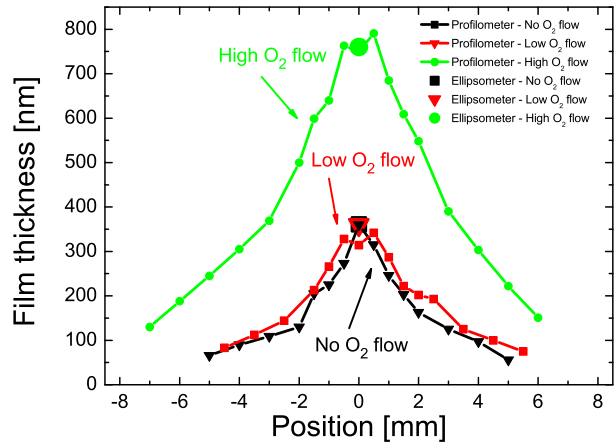}
\caption{Deposition profiles measured perpendicular to the substrate's moving direction with a profilometer and the maximum thickness measured by ellipsometry. (Conditions: Fix Parameter: 5\,slm He, 0.1\,sccm (20\,ppm) HMDSO; No O$_2$: 0\,sccm O$_2$, Profilometer $\rightarrow$ 361\,nm; Low O$_2$ flow: 2\,sccm O$_2$, Profilometer $\rightarrow$ 360\,nm; High O$_2$ flow: 15\,sccm O$_2$, Profilometer $\rightarrow$ 760\,nm) }
\label{fig:fig4}
\end{figure}

The film thickness was measured by ellipsometry and profilometry. Additionally, the profiles taken
perpendicular to the jet's moving direction in the linear drive process have been determined by
profilometry. Three selected profiles are shown in Figure~\ref{fig:fig4} together with the
maximum thickness as determined by ellipsometry. The first deposition was carried out without using
oxygen leading to a carbon rich film. For the second deposition, an oxygen flow of 2\,sccm was used
to achieve a film with a very limited carbon content and a low deposition rate. Increasing the
oxygen flow to 15\,sccm for the third deposition also leads to a very limited carbon content in the
film but to a high deposition rate. These three films were deposited under the same conditions but
varied oxygen flow. The high deposition rate at the high oxygen flow leads to a more than two times
larger film thickness compared to the low/no oxygen flow conditions. Additionally, the FWHM of the
profiles changes with the O$_2$ admixture being 3.7\,mm, 4.4\,mm and 5.6\,mm for 0, 2 and 15\,sccm
of O$_2$ respectively. The results from the ellipsometer measurements are in good agreement with
the results acquired with the profilometer. All further measurements of the film thickness shown in
this article have therefore been carried out only with ellipsometry.

\begin{figure} \centering
\includegraphics*[width=8cm]{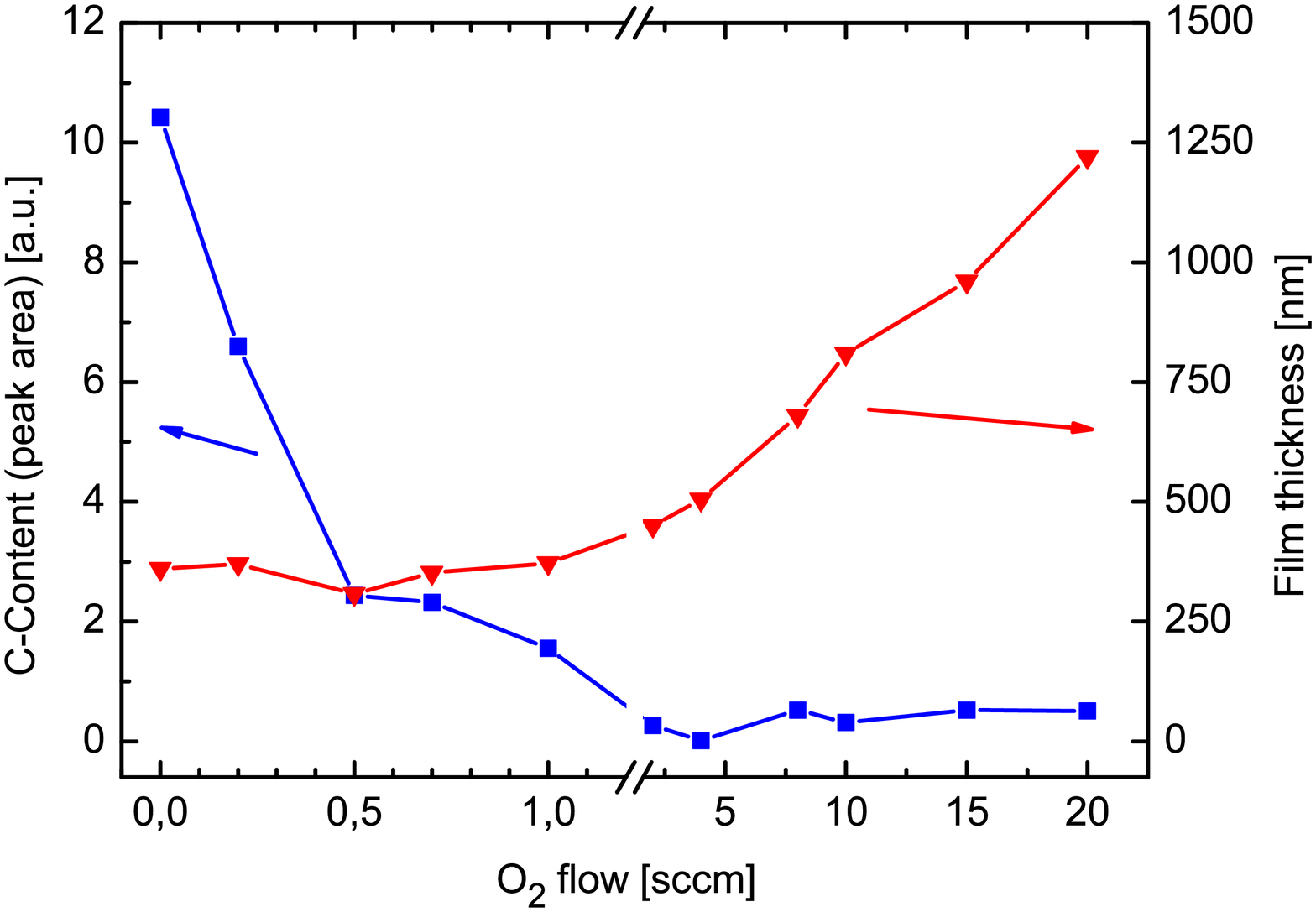}
\caption{Carbon content measured by the peak area of the carbon peak in the normalized FTIR spectrum at around 1260 wavenumbers under variation of the oxygen flow and the absolute film thickness. (Conditions: 5\,slm Helium, 0.1\,sccm HMDSO v=0.5\,mm$\cdot$min$^{-1}$)}
\label{fig:fig5}
\end{figure}

\begin{figure} \centering
\includegraphics*[width=8cm]{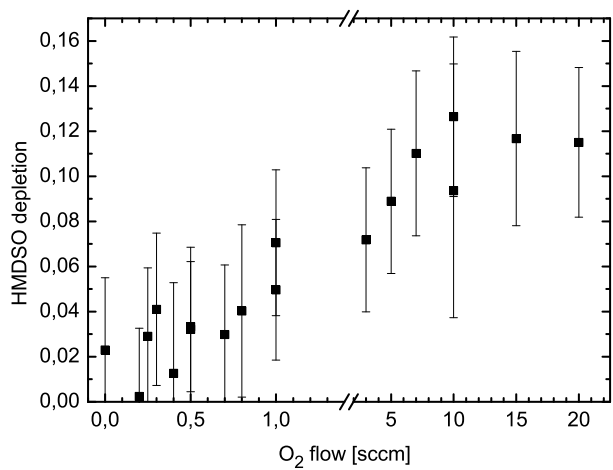}
\caption{
Depletion of HMDSO measured by molecular beam mass spectrometry. (Condition: 5\,slm He, 0.1\,sccm HMDSO, planar jet)}
\label{fig:fig6}
\end{figure}

Figure~\ref{fig:fig5} shows the carbon-content and the deposition rate as a function of the oxygen
flow. Here, the area under the 1260\,cm$^{-1}$ peak in the normalized FTIR spectrum serves as
measure for the carbon content. For oxygen flows smaller that 2\,sccm the deposition rate stays
constant and the carbon content decreases with increasing oxygen flow. For higher oxygen flows
(higher than 2\,sccm) the carbon content stays constant and the deposition rate increases.

The dissociation of the HMDSO molecule respectively the depletion of HMDSO has been measured by
molecular beam mass spectrometry. Because of the very low concentration of 20\,ppm of HMDSO in the
fed gas and a consumption of around ten percent, the measurement of concentration of around 2\,ppm
is quite challenging and very close to the detection limit of the mass spectrometer leading to
large error bars in Figure~\ref{fig:fig6}. A HMDSO depletion of around 3\% is observed with oxygen
flows below 2\,sccm and it increases with the oxygen flow above 2\,sccm. In comparison with
Figure~\ref{fig:fig5}, the HMDSO depletion and the deposition rate increase with increasing oxygen
flow.

The oxygen in the plasma seems to induce at least two mechanisms in the plasma. On one hand the removal
of carbon, where even small amounts of oxygen lead to a film with low carbon content or even a
carbon-free film. On the other hand, oxygen leads to an increase in the deposition rate. This is
due to changes in the plasma conditions and a more effective dissociation of the HMDSO molecule.

The measured profiles and measured HMDSO depletion allow a quantitative estimation of how many of the HMDSO
molecules are finally in the film. A 1\,mm long part of the film deposited within the linear drive
process with a HMDSO flow of 0.1\,sccm and an oxygen flow of 15\,sccm has a volume of $6.4 \cdot
10^{-3}$\,mm$^3$. Using the density of the SiO$_2$ film of around 2.5\,g$\cdot$cm$^3$,  1.6$\cdot
10^{17}$ SiO$_2$units are located in this volume. A 1\,mm long part of the film is deposited in 2\,min
using a velocity of 0.5\,mm$\cdot$min$^{-1}$ leading to an emission of 2\,sccm of HMDSO in two
minutes from which around 10\% are depleted for an oxygen flow of 15\,sccm, see
Figure~\ref{fig:fig6}. Assuming one HMDSO molecule ending up in one SiO$_2$ unit and 1\,sccm equal
to $2.7 \cdot 10^{19}$ molecules, around 3\% of the HMDSO ends up in the film. This means, one of three HMDSO dissociation products end up in the film under these conditions.

\begin{figure} \centering
\includegraphics*[width=8cm]{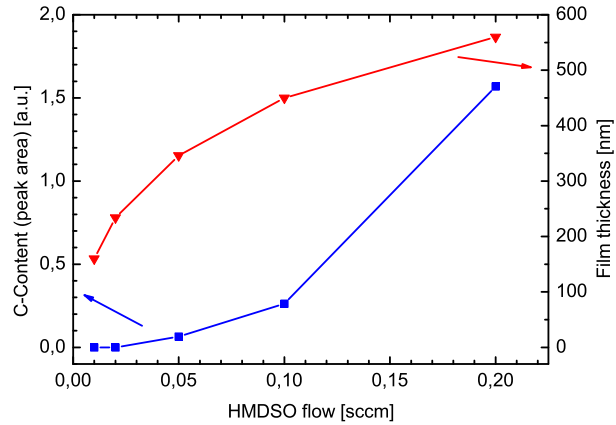}
\caption{Carbon content measured by the peak area of the carbon peak in the normalized FTIR spectrum at around 1260 wavenumbers under variation of the O$_2$ / HMDSO flow and the absolute film thickness. (Conditions: 5\,slm Helium, O$_2$ flow / HMDSO flow = 20, v=0.5\,mm$\cdot$min$^{-1}$).}
\label{fig:fig7}
\end{figure}

In the next step, the precursor flow has been decreased to check the influence on the film
properties. The results are plotted in Figure~\ref{fig:fig7}. A decreasing HMDSO flow at a constant
ratio of 20 of oxygen flow vs. HMDSO flow leads to a decrease in deposition rate and to the
disappearance of all carbon-related absorption features in the FTIR spectrum. The carbon content is
below the detection limit of the FTIR for low precursor flow conditions.

\begin{figure} \centering
\includegraphics*[width=8cm]{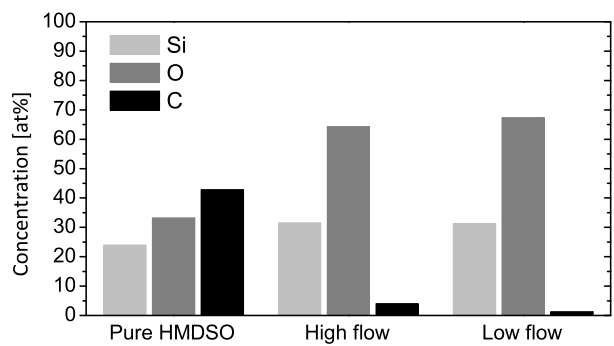}
\caption{XPS measurements (Conditions: 5\,slm He; \textit{Pure HMDSO}: 0.1\,sccm HMDSO, 0\,sccm O$_2$; \textit{High flow}: 0.1\,sccm HMDSO, 2\,sccm O$_2$;  \textit{Low flow}: 0.01\,sccm HMDSO, 0.2\,sccm O$_2$)}
\label{fig:fig8}
\end{figure}

The carbon content measurements carried out with the FTIR have been confirmed by XPS measurements.
To check in influence of exposing the probes to the ambient atmosphere, film with and without
exposure to the ambient atmosphere were measured and no difference in the atomic concentration was
observable. Every result from XPS measurements shown here was taken from films that had been expose
to the ambient atmosphere. Figure~\ref{fig:fig8} shows the atomic concentration of carbon, silicon
and oxygen in the film deposited under three different conditions. The carbon content is higher
than 40\,at\% and the O/Si ratio is around 1.4, bigger than O/Si of 0.5 in an HMDSO molecule, in a
deposition without addition of oxygen (condition \textit{Pure HMDSO}). It is stressed here that
this high ratio is not caused by a post oxidation, because the same composition is measured without
exposing the film to the ambient atmosphere. For an oxygen flow of 2\,sccm (condition \textit{High
flow}) the carbon content decreases to 4\,at\% and for a low precursor flow (condition \textit{Low
flow}) to around 1\,at\%. The O/Si ratio close to 2 is obtained for the latter two conditions.

\subsection{Planar jets: Rotating substrate experiments}

It has been shown that the important reaction for the carbon removal from the film is a surface
reaction\cite{Reuter2011}. This has been investigated with experiments on a rotating substrate,
where the simultaneous action of more jets on the same track on the rotating disk is applied and
deposition and surface treatment are separated. The deposition process has been separated
into two independent plasmas in two jets. The first jet was driven with a He/HMDSO gas mixture and
leads to a deposition of a SiO$_x$C$_y$H$_z$ film and the other one uses He/O$_2$ gases. If only
the He/HMDSO jet was applied, the film was carbon rich, when also the He/O$_2$ was applied, the
film was carbon-free as also achieved by driving a single jet with He/HMDSO/O$_2$. As already
mentioned the plasma is not in a direct contact with the surface. Charged species and short live
time metastables such as He metastables will therefore not reach the surface. The He/O$_2$ jet produces effectively O and O$_3$. This particles have a relatively long life time and are transported to the surface. This leads to the assumption that atomic oxygen is very probably responsible for the carbon etching.
It was already argued that atomic oxygen produced by the planar jet is responsible for etching of hydrogenated amorphous carbon films\cite{schneider2011}.

The rotating disc process can also be used to measure the penetration depth of reactive oxygen species formed in He/O$_2$ plasma, i.e. how many monolayers of a carbon rich
film can be deposited and afterwards treated with oxygen plasma forming a carbon-free film. It is achieved by a variation of the rotation frequency. A low frequency leads to a deposition of many
monolayers per rotation and vise versa. The number of monolayers deposited in one rotation is therefore inversely
proportional to the frequency and directly proportional to the treatment time of the oxygen plasma per
rotation.

\begin{figure} \centering
\includegraphics*[width=8cm]{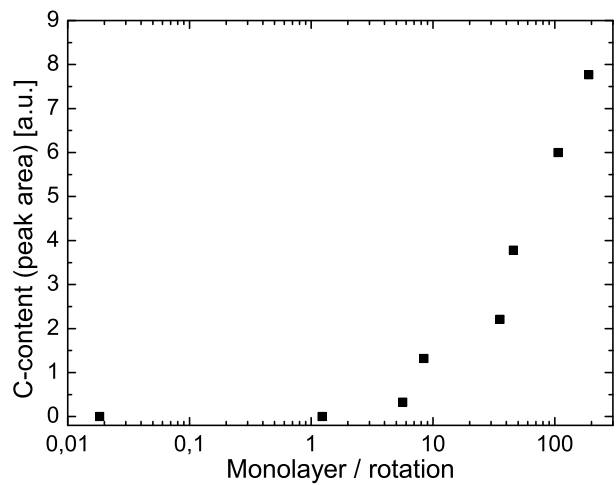}
\caption{Carbon content from depositions carried out in the rotating disc process (Conditions: 5\,slm He, 0.1\,sccm HMDSO, 2\,sccm O$_2$).}
\label{fig:fig9}
\end{figure}

For example, by doubling numbers of monolayers deposited by the HMDSO plasma, the treatment time of
the oxygen plasma also increases by a factor of two. The carbon content is
plotted over the number of monolayers deposited in one rotation in Figure~\ref{fig:fig9}.

The results from Figure~\ref{fig:fig9} show that the penetration of reactive species from the
oxygen plasma (most probably O atoms) is less than 6 monolayers. This small penetration depth is
consistent with the XPS measurements which show that exposure of the film to the ambient atmosphere
does not lead to a bulk oxidation of the deposited SiO$_x$C$_y$H$_z$ films. To calculate the number
of monolayers per rotation, a thickness of 0.3\,nm per monolayer was assumed resulting in a depth
of penetration of around 1-2\,nm.

To check the influence of the penetration time of a
carbon-rich film with helium/oxygen plasma, the helium/oxygen plasma treatment time was increased by a factor of 10. No changes in the spectra could be observed.


\subsection{Treatment of carbon rich films by an argon fed coaxial jet}

\begin{figure} \centering
\includegraphics*[width=8cm]{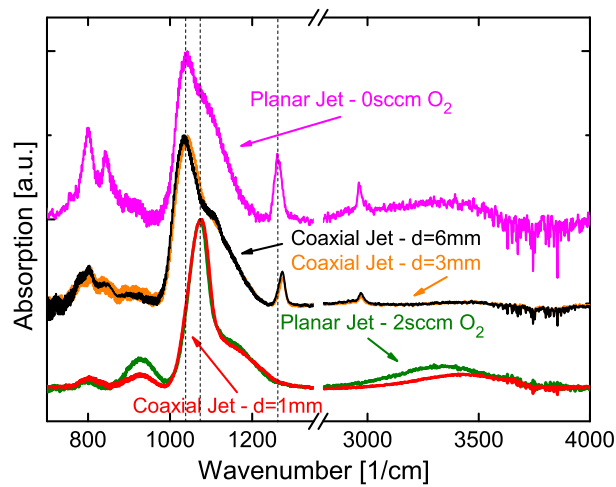}
\caption{Normalized FTIR spectra from depositions from rotating disc process. (Conditions: Fix parameter: Jet 1: planar jet, 5\,slm He, 0.1\,sccm HMDSO; Planar Jet: Jet2: planar jet, 1\,slm He ; Planar Jet + O$_2$: Jet2: planar jet, 1\,slm He, 2\,sccm O$_2$ ; Coaxial Jet, d=1mm: Jet2: Coaxial Jet 3000\,sccm / 160\,sccm Ar, distance=1mm; Coaxial Jet, d=3mm: Jet2: Coaxial Jet 3000\,sccm / 160\,sccm Ar, distance=3mm; Coaxial Jet, d=6mm: Jet2: Coaxial Jet 3000\,sccm / 160\,sccm Ar, distance=6mm). For a better visibility the FTIR spectra of conditions Planar Jet - 0sccm O$_2$, Coaxial Jet - d=3\,mm and Coaxial Jet - d=6\,mm are shifted.}
\label{fig:fig10}
\end{figure}

As already discussed, it has been shown in the past that carbon free films can be deposited even
without addition of oxygen when argon fed coaxial jet with small HMDSO flows is used
\cite{Raballand2009a}. The experiment with a rotating substrate gives us now the opportunity to
investigate, whether surface reactions can again explain the carbon removal from the growing film.
Therefore the planar jet with He/O$_2$ gas mixture in the experiment with the rotating  substrate
was replaced by the coaxial jet operated only with argon gas.  Three distances of the coaxial jet from the surface were tested: 1\,mm, where the plasma (the light emitting part of the effluent) is
in direct contact with the substrate, and 3\,mm, and 6\,mm, where the plasma is not in direct
contact with the effluent. FTIR spectra of the resulting films are shown in Figure~\ref{fig:fig10}.
The absence of the Si(CH$_3$)$_x$ bending absorption peak at around 1260\,cm$^{-1}$ in the film
treated with the coaxial jet at a distance of 1\,mm indicates that carbon can be removed from the
film. The FTIR spectrum is equal to the spectrum taken from a film deposited under the same
conditions except of replacing the coaxial jet by a planar jet driven with a He/O$_2$ mixture in
this case. On the contrary, the film is rich of carbon for distances of 3\,mm and 6\,mm, where the
surface is not in direct contact with the argon plasma. This is distinguished by the high
Si(CH$_3$)$_x$ bending absorption peak and by the shift of AS1 and AS2 to lower wave numbers. It
does not seem to play a role whether the distance is 3\,mm or 6\,mm. Compared to a deposition with
a planar jet without addition of oxygen the carbon content is lower as indicated by the height of
the Si(CH$_3$)$_x$ bending absorption peak. This experiment clearly shows that the removal of
carbon from a carbon rich SiO$_x$C$_y$H$_z$ film can be achieved by treatment with Ar plasma
without addition of oxygen and that the reaction leading to the carbon loss is a surface reaction.
It seems to be important that the plasma is in direct contact with the surface. If the film is in
contact with the plasma, ions and metastables can reach the surface and interact with the deposited
carbon rich film e.g. they can transport energy to the surface and can for example induce breaks of
the Si-C bonds.
Atomic oxygen, O$_3$ or other oxygen containing neutral reactive radicals or reactive species, originating for example from gas impurities in Ar, are most probably not responsible for the observed effect, because they would react with carbon at the surface even without direct contact of active plasma region with the surface. Additionally, the smaller content of OH in the film, visible as a smaller absorption of OH related features in the FTIR spectrum, corroborates that reactive oxygen species are not involved in the surface reactions in this case.
Currently running measurements of argon metastables performed with tunable diode laser absorption spectroscopy (TDLAS, results not shown here) indicates that argon metastable densities of $10^{13}\,cm^{-3}$ are reached at 1\,mm distance and that their density is below detection limit in front of the substrate placed at 3\,mm distance.

\section{Conclusion}

The deposition of thin SiO$_x$C$_y$H$_z$ or SiO$_x$H$_y$ films by means of planar microscale
atmospheric pressure plasma jet with helium/Hexamethyldisiloxane(HMDSO) and helium/HMDSO/O$_2$
mixtures and the role of surface reactions for the film growth have been investigated. The FTIR,
XPS, ellipsometry and mass spectrometry measurements have shown that the carbon content in and the
growth rate of the SiO$_x$C$_y$H$_z$ films is, together with the HMDSO depletion, controlled by the
admixture of oxygen into the gas mixture. The addition of oxygen has following effects: i) the
carbon content in the film is reduced and saturates at O$_2$/HMDSO ratio of 20. The deposition rate
and the depletion of HMDSO increases when more O$_2$ than this ratio is added to the gas mixture.
The lowest carbon content ($\leq\sim$1\%) is achieved at very low concentrations of HMDSO
($\sim$2\,ppm) and with addition of O$_2$. The etching of the surface, most probably by O atoms or
other reactive oxygen species (ROS) produced by the He/O$_2$ plasma, is the reaction responsible
for the carbon removal from the growing film. The penetration of ROS species has been measured in
the experiment with the rotating substrate and is less than 6\,monolayers.

The surface treatment of the SiO$_x$C$_y$H$_z$ films by coaxial jet operated with Ar and without admixture
of oxygen leads to the formation of carbon free films as well. The fact that a direct contact of
the plasma with the surface is necessary for this process to be effective and the lower content of the OH groups in the film indicates that ions
and/or excited particles (metastables) from the plasma are involved in this process.

\section{Acknowledgement}

The authors thank Norbert Grabkowski for his skillful technical assistance.
The project is supported by the German Research Foundation (DFG) in the research group FOR 1123, project C1 and by the Research Department Plasmas with Complex Interactions from the Ruhr-Universität Bochum.


\begin{thebibliography}{31}
\expandafter\ifx\csname natexlab\endcsname\relax\def\natexlab#1{#1}\fi
\expandafter\ifx\csname bibnamefont\endcsname\relax
  \def\bibnamefont#1{#1}\fi
\expandafter\ifx\csname bibfnamefont\endcsname\relax
  \def\bibfnamefont#1{#1}\fi
\expandafter\ifx\csname citenamefont\endcsname\relax
  \def\citenamefont#1{#1}\fi
\expandafter\ifx\csname url\endcsname\relax
  \def\url#1{\texttt{#1}}\fi
\expandafter\ifx\csname urlprefix\endcsname\relax\def\urlprefix{URL }\fi
\providecommand{\bibinfo}[2]{#2}
\providecommand{\eprint}[2][]{\url{#2}}

\bibitem{Magni2001}
\bibinfo{author}{\bibfnamefont{D.}~\bibnamefont{Magni}},
  \bibinfo{author}{\bibfnamefont{C.}~\bibnamefont{Deschenaux}},
  \bibinfo{author}{\bibfnamefont{C.}~\bibnamefont{Hollenstein}},
  \bibinfo{author}{\bibfnamefont{M.}~\bibnamefont{Creatore}}, \bibnamefont{and}
  \bibinfo{author}{\bibfnamefont{P.}~\bibnamefont{Fayet}},
  \textbf{\bibinfo{volume}{34}}, \bibinfo{pages}{87} (\bibinfo{year}{2001}).

\bibitem{Kim1997a}
\bibinfo{author}{\bibfnamefont{M.}~\bibnamefont{Kim}}, \bibinfo{journal}{Thin
  Solid Films} \textbf{\bibinfo{volume}{311}}, \bibinfo{pages}{157}
  (\bibinfo{year}{1997}).

\bibitem{Ricci2010a}
\bibinfo{author}{\bibfnamefont{M.}~\bibnamefont{Ricci}},
  \bibinfo{author}{\bibfnamefont{J.-L.} \bibnamefont{Dorier}},
  \bibinfo{author}{\bibfnamefont{C.}~\bibnamefont{Hollenstein}},
  \bibnamefont{and} \bibinfo{author}{\bibfnamefont{P.}~\bibnamefont{Fayet}},
  \bibinfo{journal}{Plasma Processes and Polymers} \textbf{\bibinfo{volume}{8}},
  \bibinfo{pages}{108-117} (\bibinfo{year}{2011}).

\bibitem{Leu2003}
\bibinfo{author}{\bibfnamefont{G.~F.} \bibnamefont{Leu}},
  \bibinfo{author}{\bibfnamefont{A.}~\bibnamefont{Brockhaus}},
  \bibnamefont{and} \bibinfo{author}{\bibfnamefont{J.}~\bibnamefont{Engemann}},
  \bibinfo{journal}{Science} \textbf{\bibinfo{volume}{175}},
  \bibinfo{pages}{928} (\bibinfo{year}{2003}).

\bibitem{Aumaille2000}
\bibinfo{author}{\bibfnamefont{G.}~\bibnamefont{{Aumaille, K Vall\'{e}e, A.
  Granier, A Goullet, F. Turban}}}, \bibinfo{journal}{Thin Solid Films}
  \textbf{\bibinfo{volume}{359}}, \bibinfo{pages}{188} (\bibinfo{year}{2000}).

\bibitem{Hegemann1999a}
\bibinfo{author}{\bibfnamefont{D.}~\bibnamefont{Hegemann}},
  \bibinfo{author}{\bibfnamefont{U.}~\bibnamefont{Vohrer}},
  \bibinfo{author}{\bibfnamefont{C.}~\bibnamefont{Oehr}}, \bibnamefont{and}
  \bibinfo{author}{\bibfnamefont{R.}~\bibnamefont{Riedel}},
  \bibinfo{journal}{Surface and Coatings Technology}
  \textbf{\bibinfo{volume}{116-119}}, \bibinfo{pages}{1033}
  (\bibinfo{year}{1999}).

\bibitem{Hegemann2003}
\bibinfo{author}{\bibfnamefont{D.}~\bibnamefont{Hegemann}},
  \bibinfo{author}{\bibfnamefont{H.}~\bibnamefont{Brunner}}, \bibnamefont{and}
  \bibinfo{author}{\bibfnamefont{C.}~\bibnamefont{Oehr}},
  \bibinfo{journal}{Surface and Coatings Technology}
  \textbf{\bibinfo{volume}{175}}, \bibinfo{pages}{253} (\bibinfo{year}{2003}).

\bibitem{sonnenfeld02}
\bibinfo{author}{\bibfnamefont{A.}~\bibnamefont{Sonnenfeld}},
  \bibinfo{author}{\bibfnamefont{T.~M.} \bibnamefont{Tun}},
  \bibinfo{author}{\bibfnamefont{L.}~\bibnamefont{Zakickova}},
  \bibinfo{author}{\bibfnamefont{K.}~\bibnamefont{Kozlov}},
  \bibinfo{author}{\bibfnamefont{H.-E.} \bibnamefont{Wagner}},
  \bibinfo{author}{\bibfnamefont{J.~F.} \bibnamefont{Behnke}},
  \bibnamefont{and} \bibinfo{author}{\bibfnamefont{R.}~\bibnamefont{Hippler}},
  \bibinfo{journal}{Plasmas and Polymers} \textbf{\bibinfo{volume}{6}},
  \bibinfo{pages}{237} (\bibinfo{year}{2002}).

\bibitem{Starostine07}
\bibinfo{author}{\bibfnamefont{S.}~\bibnamefont{Starostine}},
  \bibinfo{author}{\bibfnamefont{E.}~\bibnamefont{Aldea}},
  \bibinfo{author}{\bibfnamefont{H.}~\bibnamefont{de~Vries}},
  \bibinfo{author}{\bibfnamefont{M.}~\bibnamefont{Creatore}}, \bibnamefont{and}
  \bibinfo{author}{\bibfnamefont{M.~C.~M.} \bibnamefont{van~de Sanden}},
  \bibinfo{journal}{Plasma Process. Polym.} \textbf{\bibinfo{volume}{4}},
  \bibinfo{pages}{S440} (\bibinfo{year}{2007}).

\bibitem{Enache07}
\bibinfo{author}{\bibfnamefont{I.}~\bibnamefont{Enache}},
  \bibinfo{author}{\bibfnamefont{H.}~\bibnamefont{Caquineau}},
  \bibinfo{author}{\bibfnamefont{N.}~\bibnamefont{Gherardi}},
  \bibinfo{author}{\bibfnamefont{T.}~\bibnamefont{Paulmier}},
  \bibinfo{author}{\bibfnamefont{L.}~\bibnamefont{Maechler}}, \bibnamefont{and}
  \bibinfo{author}{\bibfnamefont{F.}~\bibnamefont{Massines}},
  \bibinfo{journal}{Plasma Process. Polym.} \textbf{\bibinfo{volume}{4}},
  \bibinfo{pages}{806} (\bibinfo{year}{2007}).

\bibitem{Alexandrov05}
\bibinfo{author}{\bibfnamefont{S.~E.} \bibnamefont{Alexandrov}},
  \bibinfo{author}{\bibfnamefont{N.}~\bibnamefont{McSporran}},
  \bibnamefont{and} \bibinfo{author}{\bibfnamefont{M.~L.}
  \bibnamefont{Hitchman}}, \bibinfo{journal}{Chem. Vap. Deposition}
  \textbf{\bibinfo{volume}{11}}, \bibinfo{pages}{481} (\bibinfo{year}{2005}).

\bibitem{Massines05}
\bibinfo{author}{\bibfnamefont{F.}~\bibnamefont{Massines}},
  \bibinfo{author}{\bibfnamefont{N.}~\bibnamefont{Gheradi}}, \bibnamefont{and}
  \bibinfo{author}{\bibfnamefont{A.~F.~S.} \bibnamefont{Martin}},
  \bibinfo{journal}{Surf. Coat. Technol.} \textbf{\bibinfo{volume}{200}},
  \bibinfo{pages}{1855} (\bibinfo{year}{2005}{\natexlab{a}}).

\bibitem{Premkumar2009a}
\bibinfo{author}{\bibfnamefont{P.~A.} \bibnamefont{Premkumar}},
  \bibinfo{author}{\bibfnamefont{S.~A.} \bibnamefont{Starostin}},
  \bibinfo{author}{\bibfnamefont{H.}~\bibnamefont{de~Vries}},
  \bibinfo{author}{\bibfnamefont{R.~M.~J.} \bibnamefont{Paffen}},
  \bibinfo{author}{\bibfnamefont{M.}~\bibnamefont{Creatore}},
  \bibinfo{author}{\bibfnamefont{T.~J.} \bibnamefont{Eijkemans}},
  \bibinfo{author}{\bibfnamefont{P.~M.} \bibnamefont{Koenraad}},
  \bibnamefont{and} \bibinfo{author}{\bibfnamefont{M.~C. M.~V.}
  \bibnamefont{de~Sanden}}, \bibinfo{journal}{Plasma Processes and Polymers}
  \textbf{\bibinfo{volume}{6}}, \bibinfo{pages}{693} (\bibinfo{year}{2009}).

\bibitem{Martin2004a}
\bibinfo{author}{\bibfnamefont{C.}~\bibnamefont{{Martin, S Massines, F.
  Gherardi, C. Jimenez}}}, \bibinfo{journal}{Surface and Coatings Technology}
  \textbf{\bibinfo{volume}{177-178}}, \bibinfo{pages}{693}
  (\bibinfo{year}{2004}).

\bibitem{Sawada1995}
\bibinfo{author}{\bibfnamefont{Y.}~\bibnamefont{Sawada}},
  \bibinfo{author}{\bibfnamefont{S.}~\bibnamefont{Ogawa}}, \bibnamefont{and}
  \bibinfo{author}{\bibfnamefont{M.}~\bibnamefont{Kogoma}},
  \bibinfo{journal}{Sophia} \textbf{\bibinfo{volume}{1661}}
  (\bibinfo{year}{1995}).

\bibitem{Fanelli2010b}
\bibinfo{author}{\bibfnamefont{F.}~\bibnamefont{Fanelli}},
  \bibinfo{author}{\bibfnamefont{S.}~\bibnamefont{Lovascio}},
  \bibinfo{author}{\bibfnamefont{R.}~\bibnamefont{D'Agostino}},
  \bibinfo{author}{\bibfnamefont{F.}~\bibnamefont{Arefi-Khonsari}},
  \bibnamefont{and} \bibinfo{author}{\bibfnamefont{F.}~\bibnamefont{Fracassi}},
  \bibinfo{journal}{Plasma Processes and Polymers}
  \textbf{\bibinfo{volume}{7}}, \bibinfo{pages}{535} (\bibinfo{year}{2010}).

\bibitem{Raballand2008b}
\bibinfo{author}{\bibfnamefont{V.}~\bibnamefont{Raballand}},
  \bibinfo{author}{\bibfnamefont{J.}~\bibnamefont{Benedikt}}, \bibnamefont{and}
  \bibinfo{author}{\bibfnamefont{a.}~\bibnamefont{von Keudell}},
  \bibinfo{journal}{Applied Physics Letters} \textbf{\bibinfo{volume}{92}},
  \bibinfo{pages}{091502} (\bibinfo{year}{2008}).

\bibitem{Raballand2009a}
\bibinfo{author}{\bibfnamefont{V.}~\bibnamefont{Raballand}},
  \bibinfo{author}{\bibfnamefont{J.}~\bibnamefont{Benedikt}},
  \bibinfo{author}{\bibfnamefont{S.}~\bibnamefont{Hoffmann}},
  \bibinfo{author}{\bibfnamefont{M.}~\bibnamefont{Zimmermann}},
  \bibnamefont{and} \bibinfo{author}{\bibfnamefont{A.}~\bibnamefont{von
  Keudell}}, \bibinfo{journal}{Journal of Applied Physics}
  \textbf{\bibinfo{volume}{105}}, \bibinfo{pages}{083304}
  (\bibinfo{year}{2009}).

\bibitem{Schafer2008}
\bibinfo{author}{\bibfnamefont{J.}~\bibnamefont{Sch\"{a}fer}},
  \bibinfo{author}{\bibfnamefont{R.}~\bibnamefont{Foest}},
  \bibinfo{author}{\bibfnamefont{a.}~\bibnamefont{Quade}},
  \bibinfo{author}{\bibfnamefont{a.}~\bibnamefont{Ohl}}, \bibnamefont{and}
  \bibinfo{author}{\bibfnamefont{K.-D.} \bibnamefont{Weltmann}},
  \bibinfo{journal}{Journal of Physics D: Applied Physics}
  \textbf{\bibinfo{volume}{41}}, \bibinfo{pages}{194010}
  (\bibinfo{year}{2008}).

\bibitem{Lommatzsch09}
\bibinfo{author}{\bibfnamefont{U.}~\bibnamefont{Lommatzsch}} \bibnamefont{and}
  \bibinfo{author}{\bibfnamefont{J.}~\bibnamefont{Ihde}},
  \bibinfo{journal}{Plasma Process. Polym.} \textbf{\bibinfo{volume}{6}},
  \bibinfo{pages}{642} (\bibinfo{year}{2009}).

\bibitem{Schulz-vonderGathen2007}
\bibinfo{author}{\bibfnamefont{V.}~\bibnamefont{{Schulz-von der Gathen}}},
  \bibinfo{author}{\bibfnamefont{V.}~\bibnamefont{Buck}},
  \bibinfo{author}{\bibfnamefont{T.}~\bibnamefont{Gans}},
  \bibinfo{author}{\bibfnamefont{N.}~\bibnamefont{Knake}},
  \bibinfo{author}{\bibfnamefont{K.}~\bibnamefont{Niemi}},
  \bibinfo{author}{\bibfnamefont{S.}~\bibnamefont{Reuter}},
  \bibinfo{author}{\bibfnamefont{L.}~\bibnamefont{Schaper}}, \bibnamefont{and}
  \bibinfo{author}{\bibfnamefont{J.}~\bibnamefont{Winter}},
  \bibinfo{journal}{Contr. Plasma Phys.} \textbf{\bibinfo{volume}{47}},
  \bibinfo{pages}{510} (\bibinfo{year}{2007}).

\bibitem{Knake2008}
\bibinfo{author}{\bibfnamefont{N.}~\bibnamefont{Knake}},
  \bibinfo{author}{\bibfnamefont{K.}~\bibnamefont{Niemi}},
  \bibinfo{author}{\bibfnamefont{S.}~\bibnamefont{Reuter}},
  \bibinfo{author}{\bibfnamefont{V.}~\bibnamefont{{Schulz-von der Gathen}}},
  \bibnamefont{and} \bibinfo{author}{\bibfnamefont{J.}~\bibnamefont{Winter}},
  \bibinfo{journal}{Applied Physics Letters} \textbf{\bibinfo{volume}{93}},
  \bibinfo{pages}{131503} (\bibinfo{year}{2008}).

\bibitem{Reuter2011}
\bibinfo{author}{\bibfnamefont{R.}~\bibnamefont{Reuter}},
  \bibinfo{author}{\bibfnamefont{D.}~\bibnamefont{Ellerweg}},
  \bibinfo{author}{\bibfnamefont{A.}~\bibnamefont{von Keudell}},
  \bibnamefont{and} \bibinfo{author}{\bibfnamefont{J.}~\bibnamefont{Benedikt}},
  \bibinfo{journal}{Applied Physics Letters} \textbf{\bibinfo{volume}{98}},
  \bibinfo{pages}{111502} (\bibinfo{year}{2011}).

\bibitem{Benedikt2007c}
\bibinfo{author}{\bibfnamefont{J.}~\bibnamefont{Benedikt}},
  \bibinfo{author}{\bibfnamefont{V.}~\bibnamefont{Raballand}},
  \bibinfo{author}{\bibfnamefont{a.}~\bibnamefont{Yanguas-Gil}},
  \bibinfo{author}{\bibfnamefont{K.}~\bibnamefont{Focke}}, \bibnamefont{and}
  \bibinfo{author}{\bibfnamefont{A.}~\bibnamefont{von Keudell}},
  \bibinfo{journal}{Plasma Physics and Controlled Fusion}
  \textbf{\bibinfo{volume}{49}}, \bibinfo{pages}{B419} (\bibinfo{year}{2007}).

\bibitem{Benedikt2006}
\bibinfo{author}{\bibfnamefont{J.}~\bibnamefont{Benedikt}},
  \bibinfo{author}{\bibfnamefont{K.}~\bibnamefont{Focke}},
  \bibinfo{author}{\bibfnamefont{A.}~\bibnamefont{Yanguas-Gil}},
  \bibnamefont{and} \bibinfo{author}{\bibfnamefont{A.}~\bibnamefont{von
  Keudell}}, \bibinfo{journal}{Applied Physics Letters}
  \textbf{\bibinfo{volume}{89}}, \bibinfo{pages}{251504}
  (\bibinfo{year}{2006}).

\bibitem{Benedikt2010}
\bibinfo{author}{\bibfnamefont{J.}~\bibnamefont{Benedikt}},
  \bibinfo{journal}{Journal of Physics D: Applied Physics}
  \textbf{\bibinfo{volume}{43}}, \bibinfo{pages}{043001}
  (\bibinfo{year}{2010}).

\bibitem{Pai86}
\bibinfo{author}{\bibfnamefont{P.~G.} \bibnamefont{Pai}},
  \bibinfo{author}{\bibfnamefont{S.~S.} \bibnamefont{Chao}},
  \bibinfo{author}{\bibfnamefont{Y.}~\bibnamefont{Tagaki}}, \bibnamefont{and}
  \bibinfo{author}{\bibfnamefont{G.}~\bibnamefont{{G. Lucovsky}}},
  \bibinfo{journal}{J. Vac. Sci. Technol.} \textbf{\bibinfo{volume}{4}},
  \bibinfo{pages}{689} (\bibinfo{year}{1986}).

\bibitem{Borvon02}
\bibinfo{author}{\bibfnamefont{G.}~\bibnamefont{Borvon}},
  \bibinfo{author}{\bibfnamefont{A.}~\bibnamefont{Goullet}},
  \bibinfo{author}{\bibfnamefont{A.}~\bibnamefont{Granier}}, \bibnamefont{and}
  \bibinfo{author}{\bibfnamefont{G.}~\bibnamefont{Turban}},
  \bibinfo{journal}{Plasmas Polymers} \textbf{\bibinfo{volume}{7}},
  \bibinfo{pages}{341} (\bibinfo{year}{2002}).

\bibitem{Ellerweg2010a}
\bibinfo{author}{\bibfnamefont{D.}~\bibnamefont{Ellerweg}},
  \bibinfo{author}{\bibfnamefont{J.}~\bibnamefont{Benedikt}},
  \bibinfo{author}{\bibfnamefont{a.}~\bibnamefont{von Keudell}},
  \bibinfo{author}{\bibfnamefont{N.}~\bibnamefont{Knake}}, \bibnamefont{and}
  \bibinfo{author}{\bibfnamefont{V.}~\bibnamefont{{Schulz-von der Gathen}}},
  \bibinfo{journal}{New Journal of Physics} \textbf{\bibinfo{volume}{12}},
  \bibinfo{pages}{013021} (\bibinfo{year}{2010}).

\bibitem{Benedikt2009a}
\bibinfo{author}{\bibfnamefont{J.}~\bibnamefont{Benedikt}},
  \bibinfo{author}{\bibfnamefont{D.}~\bibnamefont{Ellerweg}}, \bibnamefont{and}
  \bibinfo{author}{\bibfnamefont{a.}~\bibnamefont{von Keudell}},
  \bibinfo{journal}{The Review of scientific instruments}
  \textbf{\bibinfo{volume}{80}}, \bibinfo{pages}{055107}
  (\bibinfo{year}{2009}).

\bibitem{Massines2005a}
\bibinfo{author}{\bibfnamefont{F.}~\bibnamefont{Massines}},
  \bibinfo{author}{\bibfnamefont{N.}~\bibnamefont{Gherardi}},
  \bibinfo{author}{\bibfnamefont{a.}~\bibnamefont{Fornelli}}, \bibnamefont{and}
  \bibinfo{author}{\bibfnamefont{S.}~\bibnamefont{Martin}},
  \bibinfo{journal}{Surface and Coatings Technology}
  \textbf{\bibinfo{volume}{200}}, \bibinfo{pages}{1855}
  (\bibinfo{year}{2005}{\natexlab{b}}).

\bibitem{schneider2011}
\bibinfo{author}{\bibfnamefont{S.}~\bibnamefont{Schneider}},
  \bibinfo{author}{\bibfnamefont{J.-W.}~\bibnamefont{Lackmann}},
  \bibinfo{author}{\bibfnamefont{F.}~\bibnamefont{Naberhaus}},
  \bibinfo{author}{\bibfnamefont{J. E.}~\bibnamefont{Bandow}},
  \bibinfo{author}{\bibfnamefont{B.}~\bibnamefont{Denis}}, \bibnamefont{and}
  \bibinfo{author}{\bibfnamefont{J.}~\bibnamefont{Benedikt}},
  \bibinfo{journal}{J. Phys. D: Appl. Phys.}
  \textbf{\bibinfo{volume}{44}}, \bibinfo{pages}{295201}
  (\bibinfo{year}{2011}{\natexlab{b}}).


\end{thebibliography}
\end{document}